\documentclass[11pt]{article}
\usepackage[top=1in, bottom=1in, left=1in, right=1in]{geometry}                
\pdfoutput=1
\usepackage{graphicx}
\usepackage{amsmath}
\usepackage{amssymb}
\usepackage{epstopdf}

\usepackage{fancyheadings}

\newcommand{\ket}[1]{\left|#1\right\rangle}

\title{Emergent  Majorana Fermions and their Restricted Clifford Algebra}
\author{R. Jackiw\\
\it \small Center for Theoretical Physics\\
\it \small Massachusetts Institute of Technology\\
\it \small Cambridge, MA 02139}
\date{} 

\begin{document}
\maketitle
\thispagestyle{fancy}

\begin{abstract}
Dedicated to Ludwig Faddeev on his $80^{th}$ birthday. Ludwig exemplifies perfectly a mathematical physicist: significant contribution to mathematics (algebraic properties of integrable systems) and physics (quantum field theory). In this note I present an exercise which bridges mathematics (restricted Clifford algebra) to physics (Majorana fermions).
\end{abstract}

In quantum field theory complex quantum fields describe charged particles and anti-particles. Neutral excitations that are also self-conjugate (particle coincides with anti-particle) use real fields. Actual examples include the zero spin neutral pion. The (neutral) spin-one photon, the (hypothetical) spin-two graviton.  All these are bosons. How about self-conjugate fermions? Admittedly there is no experimental evidence for such fermions. But present day theory uses them to account for recent developments in neutrino physics. Also super symmetric partners of self-conjugate bosons must be self-conjugate fermions. 

The Majorana equation describes self-conjugate fermions in terms of a real Dirac equation. The familiar Dirac complex equation reads

\begin{center}
\includegraphics[scale=.9]{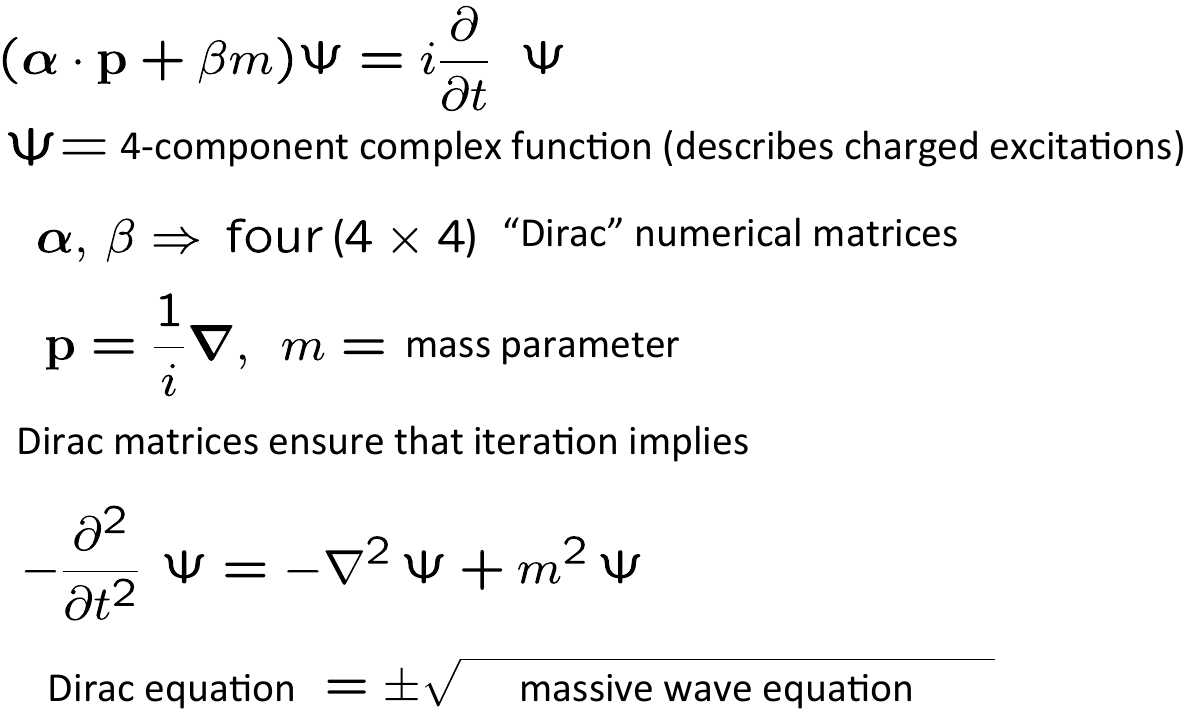}
\end{center}

The Dirac equation ``deconstructs" the massive wave equation by taking its ``square root" with matrices. But since square roots come with both signs the complex Dirac equation describes fermion particles and anti-particles. Clearly a further deconstruction is possible when particles are identified with anti-particles, {\it i.e.} when $\Psi$ is real. Upon examining the above Dirac equation, we see that $\Psi$ can be real
 if $\beta$ is imaginary, and since $\mathbf{p} =- i\,  \boldsymbol{\nabla},  \boldsymbol{\alpha}$ must be real. Such a choice of Dirac matrices defines the Majorana representation.

An explicit example of the Majorana representation for such matrices is 
\begin{eqnarray*}
\alpha^1_M = \left(\begin{array}{cc}0 & \sigma^1 \\ \sigma^1 & 0\end{array}\right)\ 
\alpha^2_M = \left(\begin{array}{cc}I & 0 \\ 0 & -I\end{array}\right)\
\alpha^3_M = \left(\begin{array}{cc}0 & \sigma^3 \\ \sigma^3 & 0\end{array}\right)\\[1ex]
\beta_M = \left(\begin{array}{cc}0 & \sigma^2 \\ \sigma^2 & 0\end{array}\right)\quad \Psi_M^\ast = \Psi_M \hspace{3em}
\end{eqnarray*}
\[
{\boldsymbol\alpha}^\ast_M = {\boldsymbol\alpha}_M , \ \ \beta^\ast_M = - \beta_M
\]
\noindent However, since the physical content is invariant against similarity  transformations, the above reality requirements can be replaced by a reality up to similarity, effected by a conjugation matrix $C$.
\[
C \boldsymbol{\alpha}^\ast \, C^{-1} = \boldsymbol{\alpha}, \ C \beta^\ast \, C^{-1}  = -\beta, \  C \Psi^\ast = \Psi
\]
For example, if we use the Weyl representation
\[
\boldsymbol{\alpha} = \left(
                                                                   \begin{array}{cc}
                                                                        \boldsymbol{\sigma} & 0 \\  
                                                                        0 & -\boldsymbol{\sigma}
                                                                        \end{array}\right)\ , \
\beta = \left(
                 \begin{array}{cc}
                         0& I \\ 
                         I & 0
                         \end{array}\right)
\]

\noindent the conjugation matrix is  $C= \left(\begin{array}{cc}0 &- i\sigma^2 \\ i\sigma^2 & 0\end{array}\right)$. (In the Majorana representation $C = I$.)

Thus the Majorana equation retains its Dirac/Weyl form, except that the condition $C \Psi^\ast =\Psi$ is imposed.
\begin{eqnarray*}
\Psi = \left(\begin{array}{c}\psi \\ \chi\end{array}\right)\qquad\
\left(
                 \begin{array}{cc}
                         \boldsymbol{\sigma} \cdot \boldsymbol{p}& m  \\ [1ex]
                         m & - \boldsymbol{\sigma} \cdot \boldsymbol{p}
                         \end{array}\right)
                         \left(\begin{array}{c}\psi \\ \chi\end{array}\right) =
                         i \frac{\partial}{\partial t}\ \left(\begin{array}{c}\psi \\ \chi\end{array}\right)\\[1ex]
                         C \, \Psi^\ast = \Psi \Rightarrow \chi = i\, \sigma^2\, \psi^\ast \hspace{2.10in}
                         \end{eqnarray*}
                         \text{The constraint leads to the two-component \textsf{Majorana Matrix Equation}.}                         
                         \begin{eqnarray*}
                         \boldsymbol{\sigma}  \cdot \boldsymbol{p} \, \psi + i \sigma^2\, m \, \psi^\ast =  i \frac{\partial}{\partial t}\ \psi\qquad (2\times 2)
\end{eqnarray*}
Note that $\psi$ mixes with $\psi^\ast$; the Majorana mass term does not preserve any quantum numbers; there is no distinction between particle and anti-particle since there are no quantum numbers to tell them apart; {\it i.e.} particle is its own anti-particle.

The field expansion for a charged Dirac field reads
 \[
   \Psi = \sum_{E>0} \left(a_E\, e^{-i Et}\, \Psi_E + b^\dagger_E\, e^{i Et}\, C\, \Psi^\ast_E\right) ,
    \]
with $a$ annihilating and $b^\dagger$ creating particles and anti-particles respectively. For the Majorana field we have
 \[
    \Psi = \sum_{E>0} \left(a_E\, e^{-i Et}\, \Psi_E + a^\dagger_E \, e^{i Et}\, C\, \Psi^\ast_E\right)  .
    \]
The anti-particle operators ($b, b^\dagger$) have disappeared.

The remarkable fact is that the condensed matter theorists have encountered essentially the same equation in a description of a super conductor in contact with a topological insulator. The relevant two dimensional Hamiltonian density governs the two-component $ \psi =
\left(\begin{array}{c}
\psi_1\\ \psi_2
\end{array}\right)$ and reads
\[
H = \psi^{\ast}\ ({\boldsymbol\sigma} \cdot \frac{1}{i}\ {\boldsymbol\nabla} -\mu)\ \psi + \frac{1}{2}\ (\triangle \psi^{\ast}\, i\, \sigma^2 \, \psi^\ast + h.c.) .
\]
 $
{\boldsymbol\sigma} = (\sigma^1, \sigma^2), \mu $ is chemical potential and $\triangle$ is the order parameter that may be constant: $\triangle = \triangle_0 = m$, or takes vortex profile; $\triangle ({\bf r}) = v (r) e^{i\theta}, v(0) = 0, v (\infty) = \triangle_0$.

The equation of motion follows: 
\[
\ i\, \partial_t \,  \psi = ({\boldsymbol \sigma} \cdot {\boldsymbol {\bf p}} -\mu)\ \psi + \triangle\, i\, \sigma^2 \, \psi^\ast  
\]

\noindent In the absence of $\mu$, and with constant $\triangle$ , the above system is a (2+1)-dimensional version of the  (3+1)-dimensional, two component Majorana equation! It governs chargeless spin $\tfrac{1}{2}$ fermions with  Majorana mass $|\triangle|$.

In the presence of a single vortex order parameter $\Delta ({\bf r}) = v (r) e^{i\theta}$ there exists a zero-energy (static) isolated mode
 \big[Rossi \& RJ {\it NPB} {\bf 190}, 681 (81); Fu \& Kane, {\it PRL} {\bf 100}, 096407 (08)\big] 
 \[
 \psi_0 = \# \left(
                  \begin{array}{l}
                  J_0 (\mu r) \exp\, \{-i\pi/4 - V(r)\}\\[1ex]
                  J_1 (\mu r) \exp\, \{i (\theta + \pi/4) - V(r)\}
                  \end{array}
                  \right)
 \]
 $\#$ real constant, $V^\prime (r) = v (r)$. The  Majorana field expansion now reads:\\

\[
 \begin{array}{ccccc}
\Psi & =\!\! &\negthickspace\!\! ................ \negthickspace\!\! &  + \ a \, \psi_0 & \\
&& {{\scriptstyle E \ne 0} \ \text{\small modes}}&& 
\end{array}
\]

\noindent where the zero mode operator $a$ satisfies
 \[
 \{a, a^\dagger\} = 1, a^\dagger = a \Rightarrow a^2 = 1/2 .
 \]
 How to realize $a$ on states? 
 There are two possibilities [Chamon, Nishida, Pi, Santos \& RJ; {\it PRB} {\bf 81}, 224515 (10)].

\begin{itemize}
\item[(i)] Two one-dimensional realizations: take vacuum state to be eigenstate of $a$, with possible eigenvalues $\pm 1/\sqrt{2}$.
\[
a\ket{0\pm} = \pm\ \frac{1}{\sqrt{2}}\ \ket{0\pm}
\]
There are two ground states $\ket{0+}$ and $\ket{0-}$. Two towers of states are constructed by repeated application of $a^\dagger_E$. No operator connects the two towers. 
Fermion parity is broken because $a$ is a fermionic operator. Like in spontaneous breaking, a   vacuum $\ket{0+}$ or $\ket{0-}$ must be chosen, and no tunneling connects to the other ground state.
\item[(ii)] One two-dimensional realization: vacuum doubly degenerate $\ket{1}, \ket{2}$, and $a$ connects the two vacua.
\[
\begin{array}{l}
a \ket{1} = \frac{1}{\sqrt{2}}\ \ket{2}\\[2ex]
a \ket{2} = \frac{1}{\sqrt{2}}\ \ket{1}
\end{array}
\]
Two towers of states are constructed by repeated application of $a^\dagger_E$. $a$ connects the towers. Fermion parity is preserved.
\end{itemize}
It seems natural to assume that fermion parity is preserved. Therefore, we adopt the second possibility, which may also be justified by considering a vortex/anti-vortex pair separated by a large distance. 

But we note that fermion parity violating realization (i) has a place in mathematical physics: Let $\mathcal{L}$ be Lagrange density for scalar kink $\oplus$ fermions.
\begin{itemize}
\item[]
	\begin{itemize}
		\item[]  
		$\mathcal{L} = \frac{1}{2}\ \partial_\mu\, \Phi\, \partial^\mu\, \Phi + \frac{\mu}{2}^2\, \Phi^2 - \frac{\lambda}{8}^2\, \Phi^4 + i \bar{\Psi}\, \gamma^\mu\, \partial_\mu\, \Psi - g\, \Phi\, \bar{\Psi} \Psi$
		\item[] $\mathcal{L}$ possesses SUSY for $g = \lambda, \Psi$ Majorana
		\end{itemize}		
	\end{itemize}
 One can prove from center anomaly in SUSY algebra that fermion parity can be absent [Losev, Shifman \& Vainshtein, {\it PLB} {\bf 522}, 327 (01)].\negthinspace\ It is an open question whether this curiosity has any relevance for condensed matter, indeed for any physical question. \big(For some speculation see [Semenoff \& Sodano, {\it EJTP} {\bf 10}, 57 (08)].\big)

How many states $\mathcal{N}$ are needed to represent $N$ vortices (in a fermion parity preserving fashion)? For $N=1$ we used two states: $\mathcal{N}=2$. For $N=2$, we have Hermitian $a_1$ and $a_2$, with $a^2_1 = a^2_2 = \frac{1}{2}$ and $a_1\, a_2 + a_2 \, a_1 = 0$. These two can be realized on the two states that are already present at $N=1$.
\begin{center}
\begin{tabular}{rrrl}
&&& $a_1 \ket{1} = \frac{1}{\sqrt{2}} \ket{2} \  \ a_1 \ket{2} = \frac{1}{\sqrt{2}} \ket{1}$\\[1.5ex]
&&&$ a_2 \ket{1} = \frac{i}{\sqrt{2}} \ket{2} \  \ a_2 \ket{2} =  \frac{-i}{\sqrt{2}} \ket{1}$
\end{tabular}
\end{center}

We can present these formulas explicitly by denoting the states by Cartesian 2-vectors, and the operators $a_i$ by Pauli matrices.
\begin{center}
\begin{tabular}{rrrl}
&&&$\ket{1} \sim\left(
\begin{array}{c}
 1 \\[-1ex]
  0
\end{array}\right) \ \ \ket{2} \sim \left(
\begin{array}{c}
 0 \\[-1ex]
  1
\end{array}\right)$\\[3.25ex]
&&& $a_1 =  \frac{\sigma^1}{\sqrt{2}} \  \ a_2 =  \frac{\sigma^2}{\sqrt{2}}$
\end{tabular}
\end{center}
These results verify the formulas  $\mathcal{N} = e^{\frac{N}{2}}$ for $N$ even and $\mathcal{N}= e^{\frac{N+1}{2}} $ for $N$ odd.

For $N=3$, we have three mode operators: $a_i, i= 1, 2, 3$. We cannot use three Pauli matrices to represent them; in particular we cannot set $a_3 = \frac{\sigma^3}{\sqrt{2}}$ because $\sigma^3$ is diagonal on the above Cartesian states and would lead to fermion parity violation. So for $N=3$ we must use $4 \times 4$ Dirac matrices and Cartesian 4-vectors; viz $\mathcal{N} = 4$.
\[
\text{states:} \ \ \ket{1} \sim
\left(
\begin{array}{c}
 1 \\[-1ex]
  0\\[-1ex]
  0\\[-1ex]
0
\end{array}
\right) \, ,
\ket{2} \sim
\left(
\begin{array}{c}
 0 \\[-1ex]
  1\\[-1ex]
  0\\[-1ex]
0
\end{array}
\right) \, , 
\ket{3} \sim
\left(
\begin{array}{c}
 0 \\[-1ex]
  0\\[-1ex]
  1\\[-1ex]
0
\end{array}
\right) \, ,
\ket{4} \sim
\left(
\begin{array}{c}
 0 \\[-1ex]
  0\\[-1ex]
  0\\[-1ex]
1
\end{array}
\right)
\]
\[
\text{operators:} \ \ \boldsymbol{\alpha}  =\left(
						\begin{array}{cc}
						0  & i {\boldsymbol \sigma}  \\
						- i {\boldsymbol \sigma}  & 0 
						\end{array}
						\right), \ \ \beta =  \left(
						\begin{array}{cc}
						 0 & I \\
						 I & 0
						\end{array}
						\right)\hspace{10em}
\] 
Choose 3 of 4 Dirac matrices $\frac{1}{\sqrt{2}} ({\boldsymbol\alpha}, \beta)$. They square to $\frac{1}{2}$, anti-commute with each other, and act on the 4 basis vectors. Thus we 
verify $\mathcal{N} = 2^{\frac{N+1}{2}} = 4 \ \text{for} \ N= 3$.

 \vtop{\hsize=6in [The matrix $\left(
                                                                               \begin{array}{cc}
							 I  &0  \\
							0 & -I 
							\end{array}
							\right)$
	also anti-commutes but cannot be used because it is diagonal, and would lead to fermion parity violation.]}

\vspace{2ex}	
 Higher N:  The pattern is now clear. We use a Clifford algebra realized by $\mathcal{N} \times \mathcal{N}$ matrices and $\mathcal{N}$-component  Cartesian vectors to represent $N$ zero-mode operators acting on states.
In selecting the members of the Clifford algebra, diagonal elements (in the Cartesian basis) must not be used, because they correspond to fermion parity violating realizations.
The fermion parity preserving realization leads to a number of states given by formulas $\mathcal{N} = 2^{\frac{N}{2}}$ and $\mathcal{N} = 2^{\frac{N+1}{2}}$ for even number and odd number vortices, respectively. The emergent algebra is a Clifford algebra, with the restriction that diagonal elements are not present [Pi \& RJ, {\it PRB} {\bf 85}, 033102 (12)].

It is an interesting open question, what role, if any, should be assigned to the fermion parity violating realizations, which arise in supersymmetry. 

These days we anticipate hearing from experimentalists that Majoranas have been found. But who will be first: condensed matter or particle physicists?
\end{document}